Correlations among superconductivity, structural instability, and band filling in $Nb_{1-x}B_2$ at the critical point x≈0.2

R. J. Xiao, K.Q. Li, H. X. Yang, G.C. Che, H.R. Zhang, C. Ma, Z.X. Zhao, and J.Q. Li*

Beijing National Laboratory for Condensed Matter Physics, Institute of Physics, Chinese Academy of Sciences, Beijing, 100080, P.R. China

We performed an extensive investigation on the correlations among superconductivity, structural instability and band filling in $Nb_{1-x}B_2$ materials. Structural measurements reveal that a notable phase transformation occurs at x≈0.2, corresponding to the Fermi level ($E_F$) in the pseudogap with the minimum total density of states (DOS) as demonstrated by the first-principles calculations. Superconductivity in $Nb_{1-x}B_2$ generally becomes visible in the Nb-deficient materials with x≥0.2. Electron energy-loss spectroscopy (EELS) measurements on B $K$-edge directly demonstrated the presence of a chemical shift arising from the structural transformation. Our systematical experimental results in combination with theoretical analysis suggest that the emergence of hole states in the σ bands plays an important role for understanding the superconductivity and structural transition in $Nb_{1-x}B_2$.

Magnesium diboride ($MgB_2$), as a phonon-mediated superconductor with $T_c \approx 39K$, has attracted considerable interest from both the theoretical and experimental viewpoints [1-6]. The superconducting transitions in the $MgB_2$ materials synthesized under a variety of conditions appear at around the limit of $T_c$ as was suggested theoretically several decades ago for BCS [3, 5, 7]. Actually, superconductivity in the metal borides with an $AlB_2$-type structure has been studied for several decades [8, 9]. In 1970, Cooper *et al.* reported the presence of superconductivity in both the Nb-B and Mo-B systems, and they also noted that the stoichiometric compounds, $NbB_2$ and $MoB_2$, in general were non-superconductors, and a few B-rich samples such as $NbB_{2.5}$ show sharp superconducting transitions [10]. Recent investigations [7-10] suggest that the Nb concentration in the $Nb_{1-x}B_2$ materials plays a significant role for the appearance of superconductivity. In the present study, we have prepared a series of samples with nominal compositions of $Nb_{1-x}B_2$ ($0<x<0.7$) under a high-pressure of 6GPa. The structural properties, superconductivity, electronic structures and the EELS spectra have been systematically analyzed along with the modification of the Nb deficiency.

Samples with nominal compositions of $Nb_{1-x}B_2$ (x=0.0, 0.1, 0.2, 0.3, 0.4, 0.5, 0.6, and 0.7) were synthesized under high pressure from Nb powder (99.9%), and B amorphous (97%). The amorphous B was preheated at 800°C for 12 hours in a vacuum to remove humidity and other volatile impurities. Mixed powers with the ratio for various x values were mounted into a sintered BN tube. The samples were heated to 1200°C in 10 min and sintered for 0.5 hour under a pressure of 6GPa using a cubic-anvil-type equipment. They were quenched to room temperature within a few seconds. The TEM investigations were performed on a Tecnai F20 (200 kV) electron microscope equipped with a post column Gatan imaging filter. The energy resolution in the EELS spectra is 0.7 eV under normal operating conditions. Under parallel illumination, the convergence angle is about 0.7 mrad and the spectrometer collection angle is ~1.0 mrad.

Theoretical calculations were performed based on the experimental structural parameters as given in fig. 1a and ref. [7]. The theoretical EELS data were simulated by the TELNES program of the WIEN2k distribution [11]. Self-consistency was carried out on a 12×12×10 mesh for $NbB_2$. The $R_{mt}K_{max}$ was set to 7.0 to determine the basis size. Both the energy and the charge density were well converged, to less than $10^{-5}$ and $10^{-4}$, respectively. For the disordered Nb-vacancy calculations, we use the SPR-KKR code [12] which accounts for the disordered atoms by using the Coherent Potential Approximation (CPA) alloy theory [13].

The crystal structures and DC susceptibilities of the $Nb_{1-x}B_2$ materials have been firstly measured by X-ray diffraction (XRD) and SQUID, respectively. The experimental results demonstrated that the $Nb_{1-x}B_2$ samples with $0<x<0.2$ occasionally contain some additional weak peaks in the XRD patterns, possibly arising from impurity phases. Measurements of DC susceptibility on the materials with $0 \leq x < 0.2$ show very weak diamagnetism at low temperatures, suggesting the existence of superconducting phase in a small fraction of a sample. Samples with $0.3 \leq x \leq 0.5$ in general show very sharp superconducting transitions just below the critical temperatures of around 8K.

Figure 1a displays the compositional dependence of the lattice parameters (*a* and *c*). It is remarkable that both the *a* and *c* lattice parameters show abrupt changes at around

x=0.2; i.e., an evident contraction within the *a-b* plane and an expansion along the *c*-axis direction. This abrupt change cannot be attributed simply to the Nb-deficiency, the alternation of electronic structure is likely to be the key reason to account for this transition as discussed in the following context. Previously, structural instability and changes of lattice parameters were reported by Yamamoto *et al*. [7], who found sudden changes of lattice parameters in a slightly Nb-deficient sample, $Nb_{0.96}B_2$. In our experiments, we carefully checked the structural properties of several series of samples prepared under different conditions, and our measurements indicate that the phase transition in general occurs at around x=0.2.

Figure 1b shows the temperature dependences of DC susceptibility for a number of $Nb_{1-x}B_2$ samples. The inset illustrates the data for 0≤x≤ 0.2, showing the presence of weak diamagnetism arising from a small fraction of superconducting phase in the materials. This weak superconductivity was considered to arise from a superconducting impurity, such as unreacted Nb-metal [7]. Large superconducting volume fraction and sharp transitions at ~7.5K were observed in most samples with x ranging from 0.3 to 0.5. The maximum critical temperature $T_c$ is observed at the temperature of 8.1K in $Nb_{0.7}B_2$ material. In fig. 1c we display the superconducting transition temperature ($T_c$) and the superconducting volume fraction (SVF) as the functions of Nb deficiency (x) obtained in our experiments. Our systematic measurements suggest that the superconductivity chiefly exists in the range of 0.2<x<0.4, in good agreement with the theoretical results obtained by Thomas Joseph *et al.* [14], who suggested the superconductivity appears progressively as x grows larger than 0.2.

In order to better understand the effects of Nb vacancies on the electronic structure of $Nb_{1-x}B_2$ materials, we have further calculated the electronic band structure and density of states (DOS) of $NbB_2$ using the augmented plane wave (APW) method by WIEN2k code [11]. Figures 2a and b show the LDA band structure and DOS for $NbB_2$, and these results are fundamentally in agreement with the results obtained in ref. [15]. The electronic states near the Fermi level ($E_F$) are mainly Nb $d$ and B $p$ orbitals. In contrast with the flat bands of the $MgB_2$ superconductor along the $\Gamma-A$ line (the $c$-axis direction), the electronic bands of $Nb_{1-x}B_2$ show up clear dispersion in the $c$-direction. This fact likely suggests a relatively stronger interplanar coupling existing in the $NbB_2$ system.

The electronic band structure and density of states of $Nb_{1-x}B_2$ with visible Nb vacancies ($0.1 \leq x \leq 0.3$) were calculated by CPA method to simulate the effects of the random distributed Nb vacancies, and fig.3 shows the DOS and band structures for $Nb_{1-x}B_2$ with x=0.1, 0.2 and 0.3. It is notable that the total DOS of these materials commonly has a sharp valley at around the $E_F$ (which was termed as a pseudogap in certain previous publications), as clearly indicated by arrows in fig. 3a. All $Nb_{1-x}B_2$ compounds possess finite $N(E_F)$ at the Fermi level, and therefore, exhibit metallic behavior. It is notable that the introduction of Nb vacancies could result in remarkable shifts of the Fermi level with significant changes of DOS near the $E_F$. The Bloch spectral function, plotted as a "smeared" band structure, is shown in fig. 3b. The most striking feature is the broadening of the σ bands at the A point along with the increasing of vacancy concentrations. In $Nb_{0.9}B_2$, the two σ bonding bands remain comparatively sharp below the Fermi level, the $E_F$ locates at ~1eV above the pseudogap and the antibonding states are partially occupied. In $Nb_{0.8}B_2$, the widths of the two σ bands are apparently enlarged especially at the A point, which makes the antibonding states shift upward and become completely unoccupied. The Fermi level shifts to the DOS minimum in the pseudogap. In $Nb_{0.7}B_2$, the broadening of the σ bands becomes notably visible due to the increase of both vacancy concentration and disorder effect. As a result, the $E_F$ shifts to about ~1eV below the pseudogap and the hole states appear in the σ bonding bands. These facts suggest that the rigid band model is not very suitable for describing the electronic band changes in present system. For instance, the valence bands in the Nb-deficient superconductors extend much greater at around A point than at the other points in the Brillouin zone. The partially occupied antibonding states in the materials with 0≤x<0.2 could result in lattice expansion within the *a-b* plane as observed in measurements of the X-ray diffraction.

In previous studies, EELS analysis has been performed on $MgB_2$ superconductor to reveal the anisotropic excitations of hole states [16, 17]. The results suggest that the pre-peak at the *K*-edge of boron in $MgB_2$ is directly related to the hole concentration [16-19]. The similar discussions on the relationship between the charge carrier and the pre-peak of the O *K*-edge have been widely analyzed in the high-$T_c$ superconducting oxides [20, 21]. Figure 4a shows the experimental B *K*-edge core-loss EELS data for several $Nb_{1-x}B_2$ samples in which the plural scattering was removed by Fourier-Log deconvolution [21]; the spectral simulation for $NbB_2$ was performed by TELNES program of the WIEN2k distribution. All the spectra contain three major features within 15eV above the edge onset: the pre-peak *a* at ~187eV, the peak *b* at ~193eV and the peak *c* at ~198eV. In order to understand the major excitations in connection each peak, we have performed a careful analysis. Actually, the primary excitations toward the unoccupied states can well explained based on the DOS of $NbB_2$ (fig.2b), in which the electronic features originating from B $\sigma(p_{xy})$ and B $\pi(p_z)$ are clearly indicated in the inset. The unoccupied states from $E_F$ to ~4eV above $E_F$ are governed mainly by the B $p_{xy}$ character around the $\Gamma$ point, which corresponds to the peak *a* in the experimental spectra. The peak *b* can be attributed to the transition from B 1*s* to the B $p_z$ states appearing mainly at about 4~8eV above the $E_F$. The peak *c* contains the excitations towards B $p_{xy}$ bands ranging from 8 to 14eV above the $E_F$. One of the striking features illustrated in fig. 4a is the intensity increase of the peak *a* along with the increase of the Nb vacancies. To quantitatively illustrate the intensity change of this peak in the B *K*-edge, all the spectra are firstly normalized to account for variations in sample thickness or acquisition times, and then, a Gaussian function is used to fit the experimental spectra as shown in Fig. 4b-d. The relative peak *a* intensity as the integral area of the first Gaussian function located at ~187 eV is shown in fig. 4e. It can be clearly recognized that the experimental peak intensities increase monotonically with the increase of Nb vacancy. This increase can be well interpreted as the hole states induced by Nb-vacancies are chiefly added to the B $p_{xy}$ band. The density of σ states therefore increases progressively near the $E_F$. It is also noted that the positions of peak *a*, *b* and *c* shift systematically about 1.1eV, 0.7eV and 0.7eV toward the high-energy direction in the superconducting samples (x>0.2) compared with the non-superconducting ones (x<0.2). This alternation is considered to arise not only from the changes of electronic structure but also from the crystal structural transformation at x≈0.2. Actually, this kind of edge shift in the EELS spectra in general is called chemical shift resulting from modifications of chemical bonds [21].

It is worth comparing the properties of electronic structures of $Nb_{1-x}B_2$ with that of the well-known $MgB_2$ superconductor. The superconductivity in $MgB_2$ is known to be driven by the σ holes in the band passing through Γ-A direction, these unoccupied states are strongly coupled with lattice vibrations [22, 23]. The DOS near the $E_F$ is mostly attributed to B $p_{xy}$ states (bands crossing the $E_F$ through the Γ-A line) and B $p_z$ states (bands crossing the $E_F$ through the Γ-M line), the Mg components near $E_F$ are actually very small. On the other hand, the band structure calculations for $Nb_{1-x}B_2$ (see fig.2a and fig.3b) reveal that the states near the $E_F$ for $NbB_2$ contains not only the B $p$ component but also a notable contribution of Nb $d$ character. The presence of Nb $d$ states near the $E_F$ could introduce covalent character in Nb-B bonds [15], and make the σ bonding bands sink below the Fermi level. The Nb deficiencies in $Nb_{1-x}B_2$ could relatively lower the Fermi level and result in additional holes in the σ states. Hence, the appearance of superconductivity in $Nb_{1-x}B_2$ materials with x>0.2 are considered fundamentally in connection with unoccupied σ states.

In summary, the structure and superconductivity of the $Nb_{1-x}B_2$ system, especially at the critical point of around x=0.2, have been systematically investigated. A number of significant phenomena have been observed along with the increase of Nb vacancies: (a) superconductivity has been clearly observed in the materials with 0.2≤x<0.5; (b) a structural transformation appears at this critical point with abrupt alterations of lattice parameters; (c) EELS measurements on the $Nb_{1-x}B_2$ materials directly revealed the chemical shifts arising from structural phase transition and visible increase of $p_{xy}$-hole states from additional Nb-vacancies. Systematic experimental results in combination with theoretical analysis suggest that the emergence of hole states in the σ bands is important for an understanding of the superconductivity and structural transition in $Nb_{1-x}B_2$ materials.

We would like to thank Prof. B.G. Liu and Prof. Y.M. Ni for their assistance and discussions. The work reported here is supported by National Natural Foundation of China and the Ministry of Science and Technology of China (973 project No:2006CB601001).

**Figure captions**

Fig 1. (a) Compositional dependence of lattice parameters, showing a remarkable structural transition at around x=0.2. (b) Temperature dependence of DC-susceptibility for $Nb_{1-x}B_2$. (c) Superconducting transition temperature $T_c$ and superconducting volume fraction (SVF) of $Nb_{1-x}B_2$ as a function of Nb deficiency (x), the SVF data are normalized with x=0.3

Fig 2. (a) The band structure for $NbB_2$. The character of B $p_{xy}$ (left panel) and B $p_z$ (right panel) are emphasized by circles proportional to their amount. (b)Total and partial density of states for $NbB_2$.

Fig. 3 (a) The total DOS (solid line), partial DOS of Nb $d$ (dashed line) and B $p$ (dotted line) calculated by CPA method in which the effects of random distributed Nb vacancy were included. (b) The CPA spectral density for $Nb_{0.9}B_2$ (left), $Nb_{0.8}B_2$ (middle) and $Nb_{0.7}B_2$ (right), plotted as a broadened band structure.

Fig 4. (a) The experimental B $K$-edge core-loss EELS for $Nb_{1-x}B_2$ samples with the plural scattering removed, and the simulated spectrum for $NbB_2$ is shown for comparison. (b)-(d) Gaussian fit (dotted line) to the experimental data (thick solid line) for x=0, 0.3 and 0.5, respectively. (e) Integral area of pre-peak *a* as a function of the vacancy concentration.

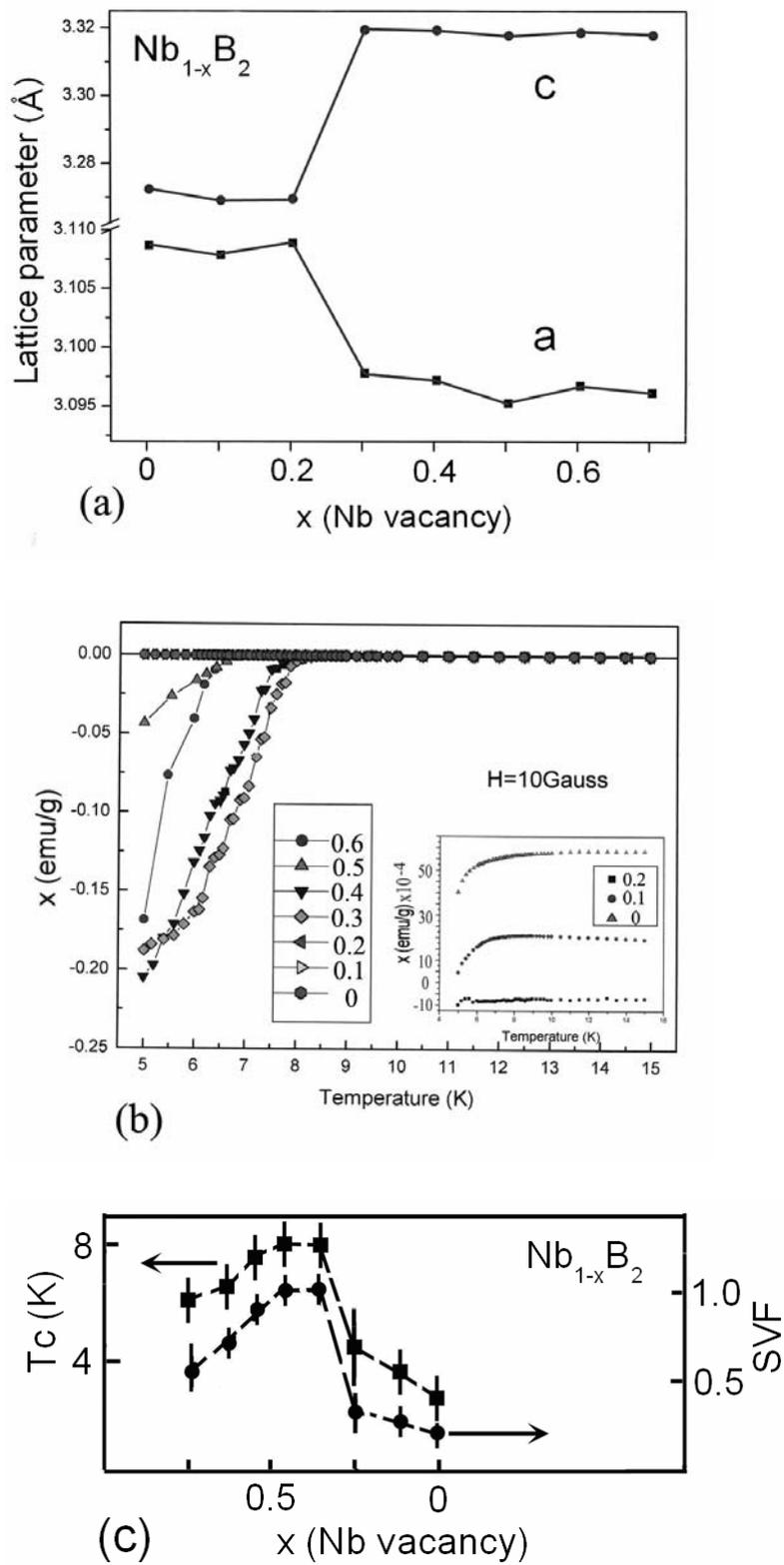

**Fig. 1**

**(a)**

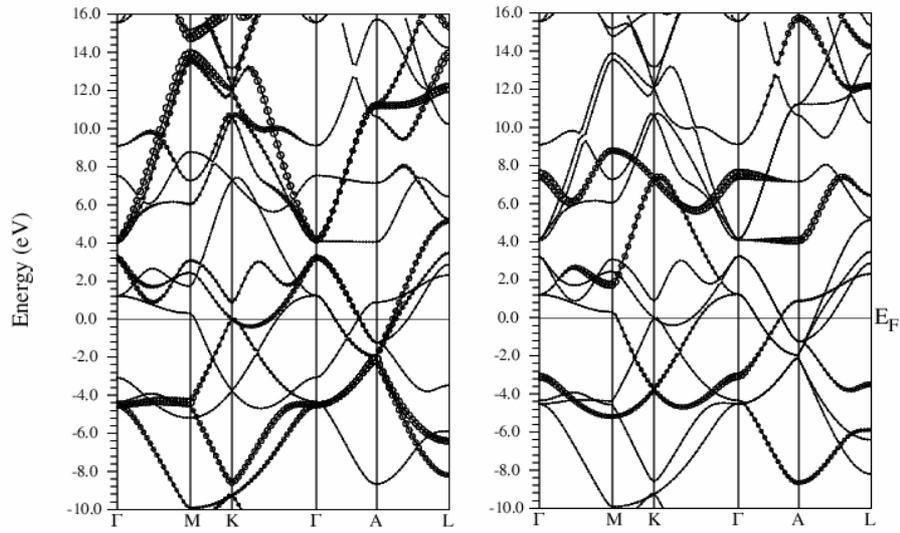

**(b)**

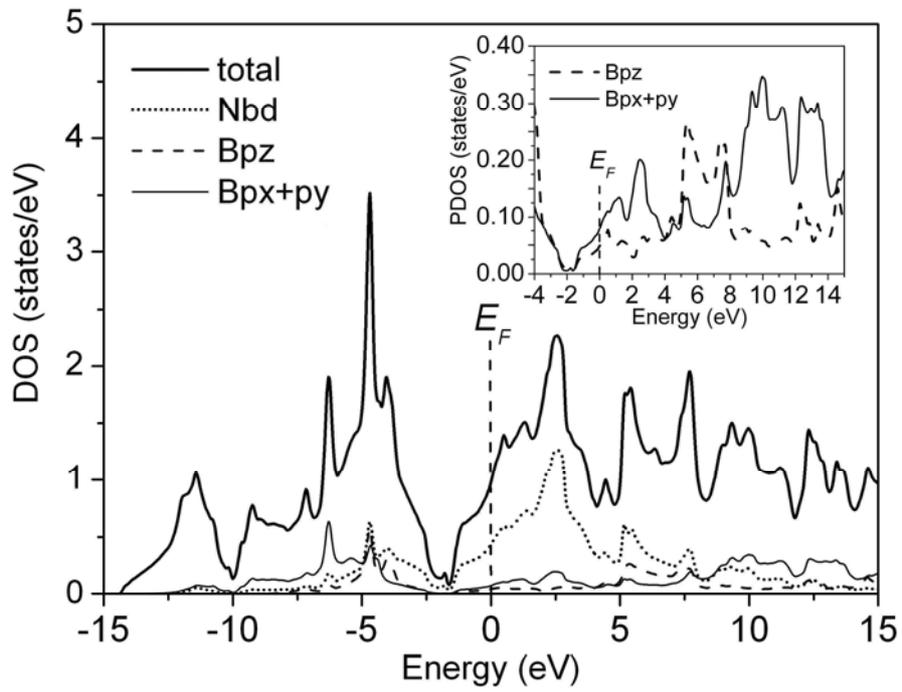

**Fig. 2**

**(a)**

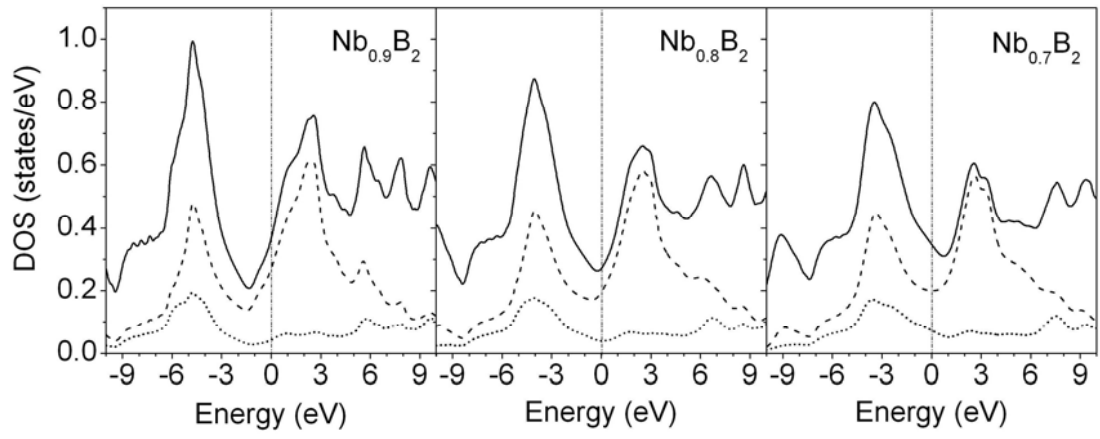

**(b)**

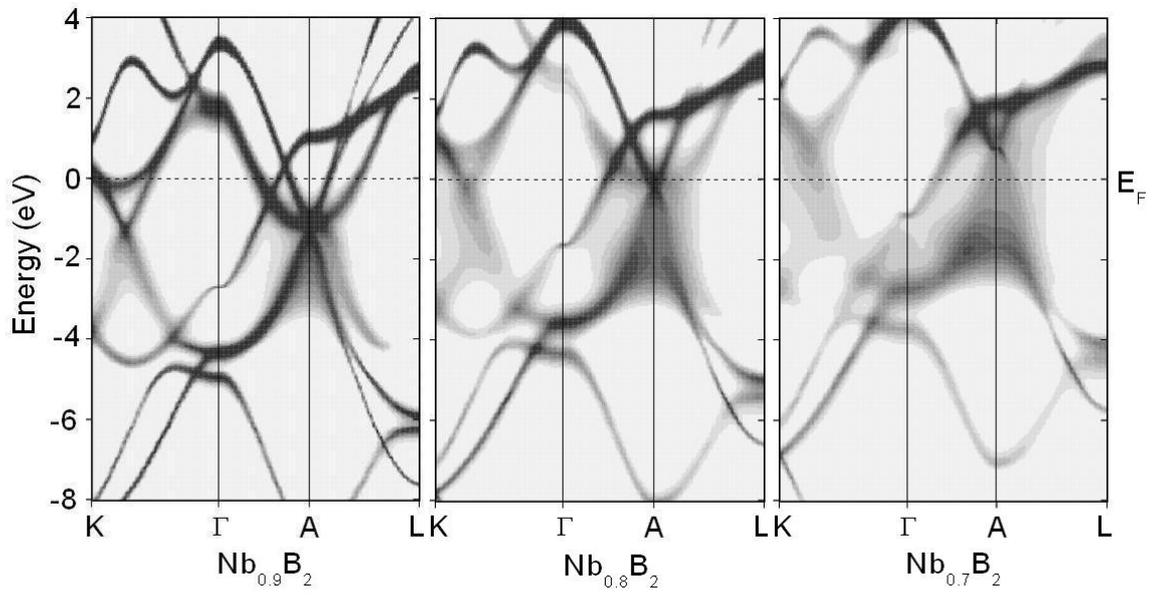

**Fig. 3**

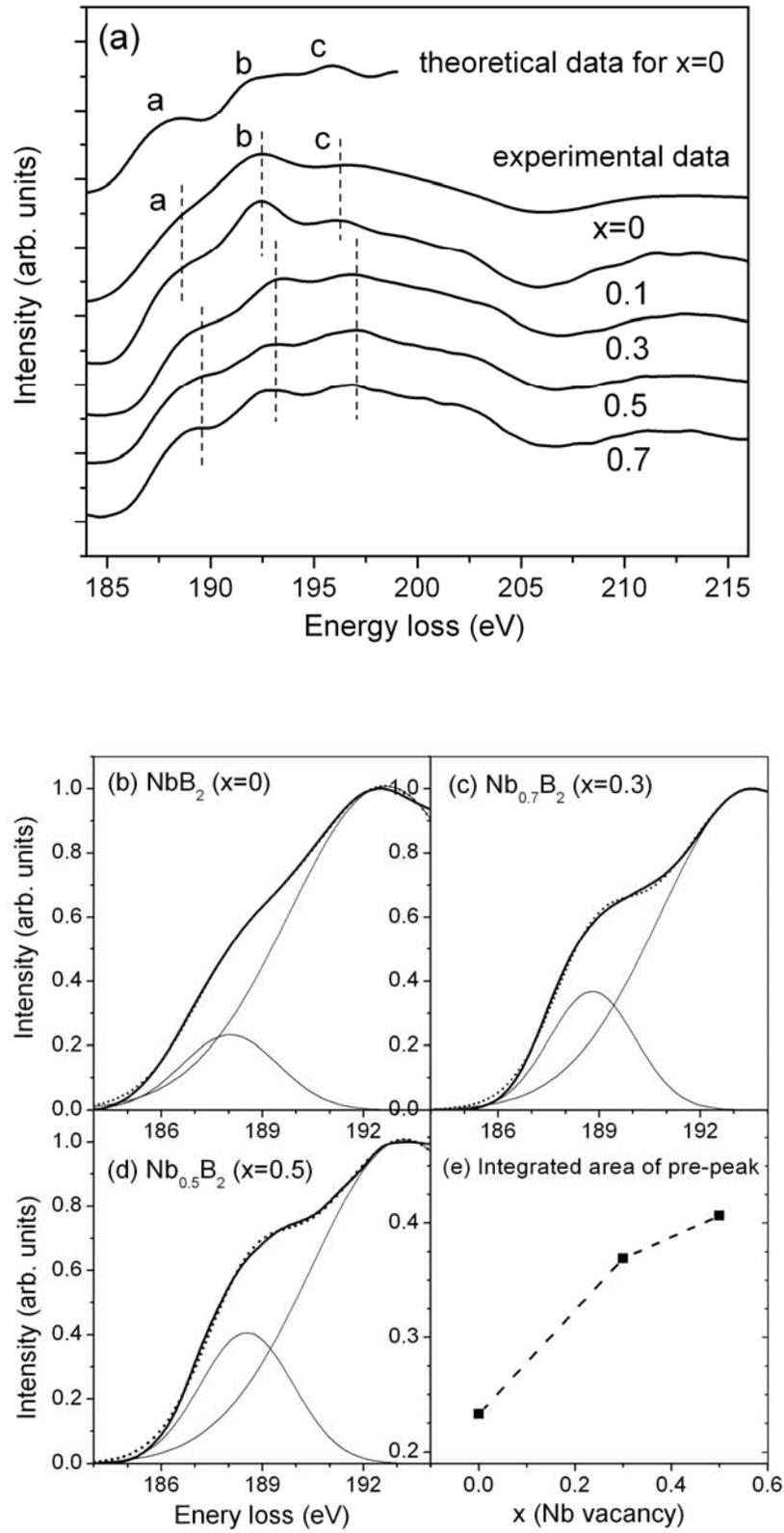

**Fig. 4**